# On-off Intermittence as a Possible Mechanism of Geomagnetic Impulse Generation

B. I. Klain, N. A. Kurazhkovskaya, and O. D. Zotov

Geophysical Observatory "Borok", Borok, Yaroslavl Region, Russia

**Abstract.** Using the many-year observations at several high-latitude observatories in the Northern and Southern Hemispheres, the regularities in distribution of magnetic impulse events (MIEs) amplitudes are studied. It is shown that the "tails" of the functions of statistical distributions of impulse amplitudes are approximated by a power function of the form $f(A) = A^{-\alpha}$, where $A$ is the impulse amplitude, and $\alpha$ is the power index laying in the range from 1 to 5. Experimental confirmation is obtained to the hypothesis suggested earlier that the impulse generation regime qualitatively corresponds to the model of on-off intermittency. It is assumed that the model of the on-off intermittency may be considered as one of the possible mechanisms of MIEs generation. According to this model, the statistical distribution of the amplitude of rare spikes is described by a power function with the power index varying from 1 to 8. An assumption is made that the value of the power index $\alpha$ may serve as a characteristic of the level of wave turbulence of the plasma in the magnetosphere in analogy to the fact that in the on-off intermittency model, the power index is a parameter of the medium in which extremely large spikes are formed. It is found that the majority of the analyzed statistical distributions of MIEs amplitudes have the index $\alpha$ exceeding 2, the fact being typical for chaotic regimes called "the strong turbulence". In some cases the index $\alpha$ is close to the unity, this fact being typical for the regimes generated in weakly turbulized medium. The results obtained make it possible to assume that: a) at higher magnetospheric latitudes (77°), the plasma turbulence degree is higher that at lower latitudes (63°); b) in the afternoon sector of the magnetosphere, the plasma is stronger turbulized than in the prenoon sector; c) in the Northern and Southern Hemispheres the magnetospheric plasma is stronger turbulized in summer and winter, respectively; d) at the growth phase of solar activity the magnetospheric plasma is stronger turbulized than at the declining phase. The dependence of the power index $\alpha$ on the solar wind speed, longitude $\phi$ of the IMF vector, and orientation of the $Bz$ component demonstrates the influence of interplanetary conditions on the high-latitude magnetosphere plasma where MIEs are formed and also the changes of the degree of magnetospheric turbulence under actions of external factors.

## 1. Introduction

During the recent decades the interest increased to studies of high-latitude magnetic impulse events (MIEs), their mean duration varying from 6 to 12 min and intensity being several tens of nT [Lanzerotti et al., 1991]. It is confirmed by numerous publications [Kataoka et al., 2003; Korotova et al., 1999; Moretto et al., 2004; Shields et al., 2003; Sibeck and Korotova, 1996]. The above mentioned specific magnetic disturbances are typical for the daytime sector of the high-latitude magnetosphere and are observed at quiet or moderate geomagnetic activity ($Kp\sim$1-3), mainly [Glassmeier et al., 1989; Lanzerotti et al., 1991; Moretto et al., 2004]. Magnetic impulses are registered from sub-auroral and auroral latitudes (62° -67°) [Glassmeier et al., 1989] to the regions of the cusp and polar cap (75° 80°) [Sibeck and Korotova, 1996]. The maximum of the impulse occurrence falls to the latitudinal range 70°-76° [Sibeck and Korotova, 1996]. The probability of impulse observations depends on geomagnetic latitude and decreases considerably at the transition from higher to lower geomagnetic latitudes. In magnetograms the impulses are often detected by a sudden pulse changes in all components of the magnetic field [Lanzerotti et al., 1991; Sibeck and Korotova, 1996].

The morphology of MIEs (spatial-temporary, spectral, and polarization characteristics) has been studied in detail in publications [Lanzerotti et al., 1991; Lin et al., 1995; Moretto et al., 2004; Sibeck and Korotova, 1996]. In some publications the relation between the impulses and solar wind and interplanetary magnetic field (IMF) parameters has been analyzed [Bering III E.A. et al., 1990; Kataoka et al., 2003; Konik et al., 1994; Moretto et al., 2004; Sibeck and Korotova, 1996]. Much enough attention has been paid in the publications to the study of the relation of MIEs to other types of magnetic disturbances and geophysical phenomena [Arnoldy et al., 1996; Konik et al., 1995; Korotova et al., 1999; Mende et al., 1990; Shields et al., 2003; Yahnin et al., 1995].

On the basis of the statistical results of magnetic impulse studies, several different mechanisms causing formation of such short-time magnetic disturbances are considered. Most often the following mechanisms are mentioned in publications: 1) reconnection processes (flux transfer events -FTE), 2) variations in the solar wind dynamic pressure, and 3) Kelvin-Helmholtz instability. According to Konik et al. [1994], the magnetic reconnection at the dayside of the magnetopause may lead to a formation a minimum of 50%-70% to a maximum of 90% of MIEs, whereas the fluctuations of the solar wind dynamic pressure are responsible for their generation only between 15% and 30% of events. The maximum 60% of MIEs is related to the Kelvin-Helmholtz instability. Kataoka et al. [2003] think that FTE and jumps in the solar wind pressure may explain ~ 30 and 20% of cases, respectively, whereas 50% of MIEs may be related to the mechanism of hot flow anomaly (HFA). Thus, currently there is no complete clearness in the problem on mechanism of magnetic impulse excitation. Moreover, each mechanism discussed in publications is able to explain generation of only some number of impulses. According to satellite data, the high-latitude magnetopause region is characterized by turbulent regions [Savin et al., 2001] what may be responsible for impulse generation. This fact assumes the presence of other mechanisms explaining generation of MIEs.

Klain and Kurazhkovskaya [2001] and Klain et al. [2004] noted earlier that magnetic impulses in appearance are similar to chaotic regimes with seldom but extremely high spikes. For the first time such regimes were described in 1990 while studying the Ginzburg-Landau equations and were called "strong turbulence" [Bartucelli et al., 1990]. However, the Ginzburg-Landau equation is a rather complicated model which does not make it possible to study completely the properties of the "strong turbulence". Simpler model, taking into account the qualitative features of the behavior of the Ginzburg-Landau equation solution and based on the principles of on-off intermittency [Platt et al., 1993], was for the first time proposed in Control [2000]. According to this paper, the exponential distribution of the interval durations between spikes and the power distribution of the spikes amplitude are the main signs of the on-off intermittency.

It follows from Control [2000] and Malinetsky and Potapov [2000] that in the case of on-off intermittence, the statistical distribution of the spikes amplitude has a long "tail" containing rare events with high amplitudes. Such events with high amplitudes can not be *apriori* excluded from the consideration because they input into the evaluation of both the mean values and dispersions. Distributions in which one can not neglect seldom spikes with high amplitudes are called in literature distributions with "heavy tails". The "tails" of such distributions are satisfactorily described by the power function of the form $f(x)= x^{-\alpha}$ at $x$ exceeding some threshold value $x_o$. It is not obligatory to have the entire distribution approximated by a power function (at all ranges of the observed values of $x$), but enough to have it approximated at $x>x_o$. The results of the numerical simulation show that in the on-off intermittence model, the "tails" of statistical distributions of the spikes amplitude are described by a power asymptotic with the power index having values $1 \le \alpha \le 8$. The distribution of the amplitude of the spikes formed in the weakly turbulized medium are approximated by the power function with the index close to the unity, whereas in the strongly turbulized medium, the value of the index is much higher and reaches 8 ("strong turbulence" regime).

Evidently, the seeming similarity of geomagnetic impulses and the regime with rare and large spikes is not a sufficient sign to identify MIEs with the on-off intermittency model. Moreover, a necessary factor is a similarity of the behavior of corresponding impulse characteristics to the known regularities of the spikes amplitude distribution and intervals between them typical for the on-off intermittency model. It should be noted, first of all, that the impulse generation is determined by quite particular conditions both in the magnetosphere itself and in the interplanetary environment. However, an unpredictable destruction of these conditions due, for example, to occurrence of external disturbances is possible. Then it is impossible to determine correctly the duration of the intervals between impulses. In other words, after an external impact, the trajectory of the system in real geophysical conditions leaves its attractor. Then the interval after which the next impulse appears would be determined not only by the time of returning to the attractor, but by the time of the system preparation to a generation of the next impulse. Due to that, the distribution of the duration of intervals between impulses will be, evidently, distorted. Therefore, the use of the statistics of distribution of intervals between impulses for identification of an on-off intermittency at generation of MIEs is incorrect. On the other hand (as numerous morphological studies have shown), the conditions occurring in the magnetosphere at impulse generation are almost identical, and that makes it possible to consider the observed impulses as belonging to the same general aggregate. Because of that, the distribution of impulse amplitudes may be used as one of important signs for identification of the MIEs regime to an on-off intermittency. The first attempts to study the character of MIEs amplitude distributions were undertaken by Klain and Kurazhkovskaya [2001] and Klain et al. [2004]. Preliminary results of these publications showed that the MIEs generation regime may be identified with the on-off intermittency model.

In this paper we continue experimental studies of regularities in distribution of geomagnetic impulse amplitudes in order to confirm the hypothesis that the on-off intermittency model may be considered as one of the mechanisms of MIEs generation.

**2. Data and Method**

We used in this paper the data of geomagnetic observations at high-latitude observatories of the Northern and Southern Hemispheres with the sweeping time of 90 mm $h^{-1}$: Heiss Island for 1974-1977, Mirny for 1981-1983 and 1985-1989, Molodezhnaya for 1981-1992, and Novolazarevskaya for 1981-1988 (all the data are from the archive of Borok observatory). Table 1 shows the codes and corrected geomagnetic coordinates of the used observatories. Analyzing the interplanetary medium conditions on against a background of which magnetic impulses were observed, we attracted the hourly data from the King Digital Catalog (the data of the WDCB, Moscow).

Magnetic impulses were identified from the analog recordings of the magnetic field using common (used in many papers) criteria. We considered as impulses sharp variations of the magnetic field having duration from 5 to 15 min and amplitude more than 10 nT. Other types of sudden changes in the magnetic field, such as sudden impulses (SI) observed on the global scale and also impulses of sudden commencements of geomagnetic storms (SSC) were out of our consideration. The total number of impulses selected at the above indicated observatories for the studied periods was, respectively 162, 75, 69, and 27 events for MIR, HIS, MOL, and NVL stations, respectively.

Table 1. List of Observatories with Codes and Corrected Geomagnetic Coordinates

| Observatory | Code | Corrected Geomagnetic Latitude, deg | Corrected Geomagnetic Longitude, deg |
|---|---|---|---|
| Mirny | MIR | -76.93 | 122.92 |
| Heiss Island | HIS | 74.8 | 144.46 |
| Molodezhnaya | MOL | -66.70 | 76.00 |
| Novolazarevskaya | NVL | -62.59 | 51.03 |

According to Control [2000], Iwasaki and Toh [1992], and Malinetsky and Potapov [2000], for identification of observed geomagnetic impulses with the regime of on—off intermittency, one has to show that the "tails" of distribution functions of MIEs amplitudes may be approximated by a power function $f(A) = A^{-\alpha}$ ($\alpha$ being the power index). The procedure of processing of the initial experimental material included the following stages.

1. Initially, the instant amplitudes of the magnetic impulses $H(t)$ and $D(t)$ for the horizontal ($H$) and azimuthal ($D$) components of the magnetic field, respectively, were evaluated. The amplitude of each impulse $A(t)$ was found from the expression: $A(t) = \sqrt{H^2(t) + D^2(t)}$. The maximum value $A(t)$ of each impulse was used for the analysis.

2. Then the entire volume of the sample of impulse amplitudes was split into intervals equal to 10 nT, the number of pulses ($n_i$) falling into the $ith$ interval was counted, and the $N = N(A)$ histograms were drawn.

3. Using the initial distributions of MIEs amplitudes, we calculated the distribution functions of the amplitude ($P$) by the formula:

$$P(A > A_m) = \int_A^\infty W(A) dA,$$

where $W(A) = N(A) / \sum_{i=1}^{k} n_i$, $k$ is the number of intervals, and $A_m$ is the least value of the impulse amplitude in the sample.

4. Some threshold value of the impulse amplitude ($A_0$) was chosen, and beginning from this value an approximation of the experimental data by a theoretical curve was performed. Since there is no formal approach making it possible to determine the threshold value $A_0$, we took as the $A_0$ approximation the point in which the nonlinear behavior of the function $\ln P = F(\ln A)$ switches to the linear behavior. The inflection point in the function $P(A)$ graph corresponds to the above indicated point. Further, the $P=P(A > A_0)$ values were normalized to the maximum value $P_{\max}$ of the $P(A > A_0)$ function. To obtain the approximating curve and confidence intervals, we used the Levenberg - Marquardt robust method [Dennis and Schnabel, 1988; Huber, 1984] realized in the MathLab packet. At the concluding stage of the study, the "tails" ($P=P(A > A_0)$) of the accumulated histograms of magnetic pulse amplitudes were approximated by the power functions, and the value of the power index $\alpha$ was determined. Below, discussing distributions of magnetic impulse magnitudes, we will mean only the distribution "tails", that is, analyze only impulses with amplitudes $A > A_0$.

5. We considered the power index $\alpha$ as the main characteristic of the MIEs amplitude distributions. To estimate the reliability degree of the obtained power indices, we used the following statistical characteristics: the standard deviation and correlation coefficient between the theoretical curve and experimental data. The upper and lower boundaries of the confidence intervals of "tails" of the distributions at the 95% level were also determined.

## 3. Results

1. Following the procedure described above we studied the amplitude distribution for the entire sampling of magnetic impulses detected at each of four observatories. Figure 1 shows the magnetic impulse amplitude distributions observed at MIR, HIS, MOL, and NVL. In the dominating number of cases, the impulse amplitude corresponded to the 10-70 nT range. At higher latitudes (MIR, HIS), the amplitude of some impulses exceeded 150-200 nT, whereas at low latitudes (MOL) it was below 100-150 nT.

The magnetic impulses observed at NVL should be specially noted. The MIEs amplitude level at NVL is much lower (10-60 nT) that at other observatories. One can see that the amplitude distributions almost at all observatories have the "tails", that is, there is relative small number of impulses with high amplitudes (see Figure 1).

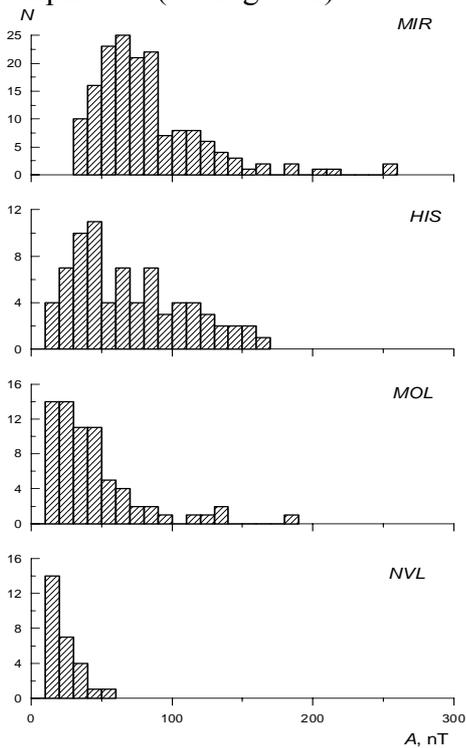

Figure 1. Distribution of amplitudes (*A*) of magnetic impulses according to the data of Mirny (MIR), Heiss Island (HIS), Molodezhnaya (MOL), and Novolazarevskaya (NVL) stations. N is the number of impulses.

Figure 2 shows the "tails" of the MIEs amplitude distribution functions at MIR, HIS, MOL, and NVL (circles), starting from some threshold value $A_0$. The value of $A_0$ for each impulse amplitude distribution was determined individually. It is worth noting that the value of $A_0$ increases with an increase of the geomagnetic latitude of observations. It should be also noted that at NVL the entire sampling of the impulses was analyzed. Solid curve shows the approximation of the experimental data by a power function. Dashed lines show the boundary of the confidential intervals. The estimation of the reliability of the obtained values of the power index showed that at the confidential probability of 0.95 the boundaries of the confidential intervals were not more than ±0.25 of the $\alpha$ value.

The standard deviation of $\alpha$ determination for all impulse amplitude distributions did not exceed the value of 0.08. The range of variations of the correlation coefficient lied within 0.94-0.99. The value of the index $\alpha$ is different at all observatories.

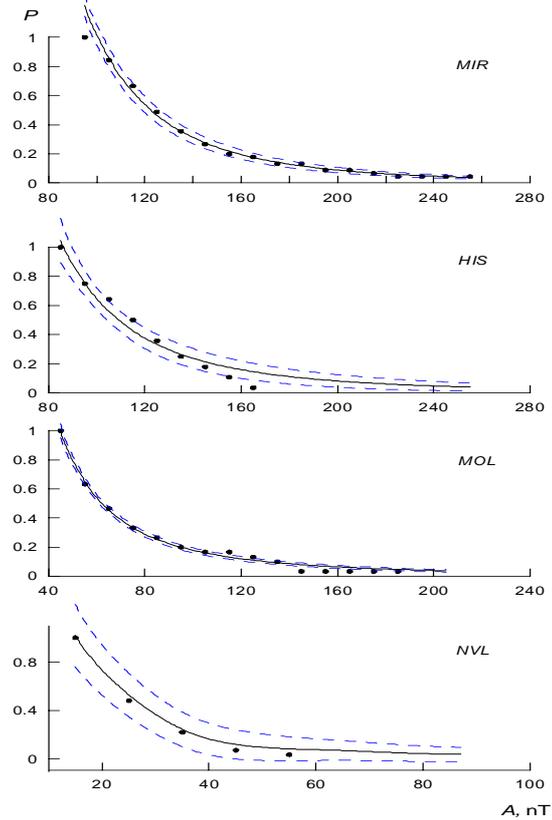

Figure 2. Approximation of the "tails" of the distribution functions of impulse amplitudes *P* (circles) by a power function (solid curve) according to the data of MIR, HIS, MOL, and NVL observatories. Dashed lines show the boundaries of confidential intervals.

There is a clear tendency of the increase in $\alpha$ value (from 1.8 at NVL to 3.5 at MIR) with an increase of the geomagnetic latitude of impulse observations (Figure 3).

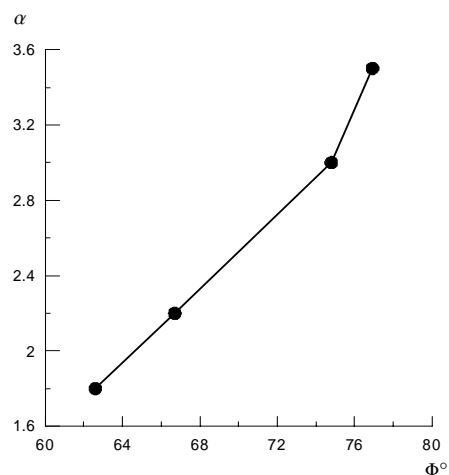

Figure 3. Dependence of the $\alpha$ index on geomagnetic latitude of observation.

2. Presented below are the results of approximation of impulse amplitude distributions and estimates of the $\alpha$ dependence on various factors. In particular, the dependence of $\alpha$ on local time, season, solar activity phase, and interplanetary conditions was studied. It should be noted here that since further all the analyzed impulses at each observatory were split into groups, the volume of the sampling decreased. The latter fact led to the increase in the standard deviation, decrease in the value of the correlation coefficient, and spreading of the confidential intervals. However, the evaluation of the errors in the index $\alpha$ for each impulse sampling taken separately showed that the errors lie within acceptable limits. For example, the standard deviation varies within 0.1-0.5, correlation coefficient was 0.79-0.99, and at a confidential probability of 95% the boundaries of the confidence intervals were spread not more than up to 0.30 of the $\alpha$ value.

The analysis of the dependence of the impulse amplitude distribution on local time showed that at all observatories there is observed a tendency to a decrease in the impulse amplitude from the prenoon to afternoon hours. Similar fact was earlier noted by other authors [Kataoka et al., 2003; Konik et al., 1995; Sibeck and Korotova, 1996]. It follows from the analysis of our data that the level of the impulse amplitudes observed in the prenoon sector is by a factor of about 1.5-2 higher than the level of impulse amplitudes registered after 1200 MLT. This conclusion is true for MIR, HIS, and MOL observatories. No such tendency is found at NVL: the average values of amplitudes in the prenoon and afternoon time are nearly the same.

Table 2. Values of the $\alpha$ Index at Various Geomagnetic Latitudes in the Prenoon and Afternoon Sectors of the Magnetosphere

| Observatory | Before 1200 MLT | After 1200 MLT |
|---|---|---|
| MIR | 3.0 | 3.8 |
| HIS | 2.7 | 5.1 |
| MOL | 1.8 | 2.5 |
| NVL | 1.4 | 1.9 |

The magnetic impulses at each observatory were split into two groups: the impulses observed before and after the local noon. Table 2 shows the values of the index $\alpha$ obtained from the MIEs observations before and after the local noon. One can see that, when the observatory is located in the prenoon sector of the magnetosphere (before 1200 MLT), the value of the $\alpha$ index is less than when the observatory is located in the afternoon sector (after 1200 MLT). This regularity is observed at the observatories located in the both hemispheres. It follows from Table 2 that the power index depends on the position of the observatory relative to the noon meridian.

3. To analyze the dependence of the power index on season, the data of observations of magnetic impulses at MIR and HIS were used. The data were split into four groups according to the commonly accepted in geophysics separation of all months of a year to seasons (around the winter and summer solstices and around the spring and fall equinoxes). Figure 4 shows the dependence of $\alpha$ on the season in the Northern and Southern Hemispheres.

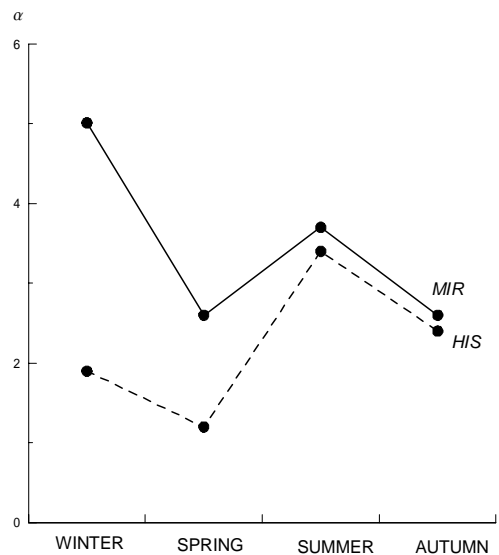

Figure 4. Dependence of the $\alpha$ index on season at MIR and HIS.

For the sake of comparison, the data of MIR and HIS are reduced to the same seasons. In the Southern Hemisphere (MIR) the value of $\alpha$ reaches its highest values in the winter season ($\alpha = 5.01$), whereas in the Northern Hemisphere (HIS) $\alpha$ takes relatively high values in summer ($\alpha = 3.40$). In the equinoxes the $\alpha$ value at both observatories is less than in the winter and summer seasons. It is worth noting that the tendency of $\alpha$ variations in the both hemispheres is identical. This experimental fact manifests the influence of the season on the MIEs amplitude behavior in both hemispheres of the magnetosphere.

4. Using the data available for two observatories (MIR and MOL), the distributions of the impulse amplitudes were analyzed and the dependence of the value of the index $\alpha$ on solar activity phase was estimated. All the cases of MIEs observations at MIR and MOL were split into two groups. The first group included the impulses observed at the decay phase of the 21st cycle of solar activity (1981-1985). The second group contained the impulses observed at the rise phase of the 22nd cycle (1986-1989). The analysis showed that at MIR, the value of $\alpha$ at the decay phase was 2.3 and increased up to 3.2 at the rise phase. Similar tendency of $\alpha$ growth from 2.0 to 2.8 depending on the phase of solar activity cycle was observed at MOL.

5. Now we come do description of the results of studies of the impulse amplitude distribution in various geophysical conditions. First, we describe briefly the interplanetary condition characteristics on against a background of which MIEs were observed. The distributions of hourly values of the solar wind velocity $V$, plasma concentration $n$, modulus of the $B$ field, dynamical pressure of the solar wind $\rho V^2$, IMF components $Bx$, $By$, and $Bz$, and latitude $\theta$ and longitude $\phi$ of the IMF vector in the solar-ecliptical coordinate system were analyzed for the cases of MIEs observations at MIR, MOL, and HIS. Since there is not enough data in the King's Catalog for the cases of MIEs observations at NVL, the corresponding analysis of the solar wind and IMF parameters was not performed.

The analysis of the interplanetary medium state (according to the hourly data on the parameters of the solar wind plasma and IMF from the King's Catalog) during observations of the magnetic impulses showed that MIEs occur both at quiet ($V < 400$ km s$^{-1}$, 30% of events) and disturbed ($V > 400$ km s$^{-1}$, 70% of events) solar wind. In the dominating number of cases, MIEs were registered at relatively low values of the solar wind density ($n \approx 4$-$6$ cm$^{-3}$) and IMF modulus ($B \approx 4$-$6$ nT). During observations of impulses, the value of the solar wind dynamic pressure ($\rho V^2$) was mainly (1-3)·10$^{-8}$ dyne/ cm$^2$. In the majority of cases, MIEs were observed at $Bx>0$, $By<0$, and $Bz>0$. The longitude $\phi$ was mainly distributed within the intervals 120°-180° and 300°-360°, this fact manifesting a radial orientation of IMF during impulse observations. In 60 % of cases MIEs were observed at the sunward direction of the IMF vector and in 40 % of cases MIEs were observed at the anti-sunward direction in the ecliptic plane. Magnetic impulses occurred mainly at the IMF orientation angle $\theta$ which lies within the interval from -30° to 30°. Thus the interplanetary medium conditions under which magnetic impulses were registered at MIR, MOL, and HIS are identical to the conditions of MIEs observations obtained according to the data of other observatories [Bering III et al., 1990; Kataoka et al., 2003; Konik et al., 1994].

Taking into account the interplanetary conditions under which MIEs were observed, we analyzed the distribution of impulse amplitudes as functions of the solar wind speed, angle $\phi$, and direction of the vertical component of IMF and obtained the following values of the power index. Table 3 show the values of the $\alpha$ index for two observatories of the Southern Hemisphere (MIR and MOL) and one observatory (HIS) of the Northern Hemisphere.

Table 3. Values of the $\alpha$ Index at Various Geomagnetic Latitudes at Various Parameters of the Interplanetary Medium

| Obs. | $V<400$ km s$^{-1}$ | $V>400$ km s$^{-1}$ | $\phi\sim(120$--$180)°$ | $\phi\sim(300$--$360)°$ | $Bz>0$ | $Bz<0$ |
|---|---|---|---|---|---|---|
| MIR | 5.0 | 3.5 | 1.9 | 2.4 | 3.0 | 2.4 |
| MOL | 1.6 | 1.1 | 1.0 | 1.2 | 0.9 | 0.8 |
| HIS | 1.3 | 3.3 | 3.4 | 2.1 | 1.9 | 2.3 |

One can see in Table 3 that the value of $\alpha$ varies considerably depending on the interplanetary medium parameters. In the speed range, all the MIEs events were split into two groups: impulses observed at $V < 400$ km s$^{-1}$ and at $V > 400$ km s$^{-1}$. In the Southern Hemisphere the values of the $\alpha$ index at $V < 400$ km s$^{-1}$ is higher than at $V > 400$ km s$^{-1}$. The opposite situation is observed in the Northern Hemisphere.

We estimated the value of the $\alpha$ index for MIEs corresponding to the negative (the anti-sunward direction of the IMF vector, $\phi \approx 120°$-$180°$) and positive (sunward direction, $\phi \approx 300°$-$360°$) IMF sectors. Here also an asymmetric behavior of $\alpha$ in two hemispheres of the magnetosphere is detected. For example, at MIR and MOL, $\alpha$ is higher at the sunward direction of the IMF vector than at the anti-sunward direction, whereas the opposite tendency is observed at HIS. Similar regularities are detected in the behavior of $\alpha$ at different directions of the IMF vertical components for the impulses registered in different hemispheres (Table 3). At the northward direction of IMF, the value of $\alpha$ at MIR and MOL is higher than at the southward orientation of IMF, whereas at HIS *vice versa* $\alpha$ is higher at $Bz<0$. Thus, considering the characteristics of the interplanetary medium for the cases of MIEs observations, an asymmetry in the $\alpha$ index behavior in two hemispheres of the magnetosphere is noted.

**4. Discussion**

It is known that geomagnetic impulses have properties characteristic for Alfven disturbances [Lanzerotti, 1989]. The main equations describing Alfven waves in plasma were presented by Mio et al. [1976] and Mjolhus [1976]. One can understand qualitatively generation in plasma of structures (impulses), considering the behavior of Alfven waves propagating along constant external field. In this case, the nonlinear Shroedinger equation [Zakharov et al., 1988]:

$$i\psi_t + \psi_{xx} + R(|\psi|^2)\psi = 0 \quad (1)$$

may serve as an acceptable model. This equation has a solution of a soliton type. If the soliton is unstable, one can hardly expect appearance of structures in turbulent regimes [Zakharov et al., 1988]. In the case of soliton stability in the turbulized plasma described by equation (1), the main input into turbulence development should be provided by solitons. Numerical calculations confirm this point of view. For the power nonlinearity ($R(u) = u^b, b \leq 2$), the calculations performed by Zakharov et al., [1988] showed that development of a modulation instability leads to a formation of a soliton grating. As a result of the interaction of the solitons with each other, the grating period increases and pumping of the energy of more weak solitons into more intense ones occurs. Finally, this leads to a formation of single impulses. Later studies showed that two-dimensional and three-dimensional generalizations of equation (1) (equations of Ginzburg—Landau, or Kuramoto—Tsuzuki) as well as equation (1) with the power nonlinearity $b > 4/d$ (where d is the space dimension) describe development of "strong turbulence" in plasma [Iwasaki and Toh, 1992]. As it has been mentioned above, the statistical distribution of peaks is characterized by a power asymptotic: $f(A) = A^{-\alpha}$. We took this very fact as the hypothesis at the study of the impulse amplitude distribution.

Simpler model taking into account the principal features of equation (1) is based on the principles of realization of on--off intermittency [Malinetsky and Potapov, 2000]. The main conclusion of this work is that the density of probability of impulse amplitudes (in the same way as in Iwasaki and Toh [1992]) has a power form. The analysis of the dependence of the power index on parameters shows that possible values of $\alpha$ lie within $1 \leq \alpha \leq 8$.

The study of distributions of magnetic impulse amplitudes observed at different geomagnetic latitudes showed that the "tails" of the obtained distributions are approximated by a power function with the power index varying (depending on various factors) from 1 to 5.0. Thus, one can conclude that the behavior of magnetic impulses qualitatively corresponds to the on—off intermittency model. Depending on local time, season, solar activity cycle phase, geomagnetic latitude of the observation, and solar wind and IMF parameters, the power index may be close to the unity (see Table 1 and Table 2), this value being typical for regimes in weakly-turbulized medium. The majority of analyzed statistical distributions of the MIEs amplitudes have the power index $\alpha$ exceeding 2 (see Table 1 and Table 2 and Figure 3 and Figure 4), this value being more typical for chaotic regimes ("strong turbulence") [Malinetsky and Potapov, 2000].

All the above said makes it possible to assume that the value of $\alpha$ may be considered as a characteristic of the turbulence level in the magnetospheric plasma in the same way as in the on—off intermittency, the power index is a parameter of the medium where exclusively large spikes are formed. Apparently, the value of the $\alpha$ index manifests the state of the magnetospheric plasma depending on local time (Table 2), season (Figure 4), geomagnetic latitude (Figure 3), solar activity cycle phase, and parameters of the interplanetary medium(Table 3). For example, lower values of $\alpha$ in the prenoon sector of the magnetosphere as compared to the afternoon sector, evidently manifest different turbulence degree of the plasma at the prenoon and afternoon flanks of the magnetosphere. Using our evaluations of $\alpha$, one can conclude that plasma is more turbulized in the afternoon sector of the magnetosphere.

The results presented in Figure 3 may be interpreted as changes in the plasma turbulence level in the magnetosphere depending on the geomagnetic latitude. Since the value of the $\alpha$ index increases at an increase of the latitude of observation, one can assume that at higher latitudes the plasma is stronger turbulized. The asymmetric behavior of the $\alpha$ index obtained from the data of impulse observations (for example, at MIR and HIS) manifests an asymmetry in the degree of plasma turbulence in the opposite hemispheres of the magnetosphere depending on the season (Figure 4). For example, the magnetospheric plasma is more turbulized in the summer season and in the season of the winter solstice in the Northern and Southern Hemispheres, respectively.

On the basis of the $\alpha$ value estimates at different phases of the solar activity cycle, one can assume that the turbulence degree of the magnetospheric plasma is higher at the rise phase of solar activity than at the decay phase.

The data presented in Table 3, show that depending on various parameters of the solar wind and IMF, the $\alpha$ index varies and so varies the state of the plasma in the magnetosphere. For example, in the Northern Hemisphere the magnetospheric plasma is more turbulized at the solar wind speed > 400 km s$^{-1}$ and at the IMF vector directed anti-sunward, whereas in the Southern Hemisphere *vise versa* the plasma is stronger turbulized at $V$ < 400 km s$^{-1}$ and the IMF vector directed sunward. The direction of the IMF vertical component also impacts the $\alpha$ value in a different way in the two hemispheres of the magnetosphere. According to the $\alpha$ estimates, the plasma is stronger turbolized at the southward and northward directions of the $Bz$ component of IMF in the Northern and Southern Hemispheres, respectively. In our opinion, the $\alpha$ dependence on the interplanetary conditions may be caused by the fact that, under the action of external factors, the medium in which magnetic impulses are generated changes: actually the plasma of the high-latitude magnetosphere undergoes alterations. Naturally, this influences the character of MIEs amplitudes distribution. Thus, the value of the power index $\alpha$ makes it possible to estimate at a qualitative level the turbulence degree of the Earth magnetosphere plasma.

## 5. Conclusion

Using the data of observations at a series of high-latitude observatories, we studied the character of distribution of magnetic impulse amplitudes observed in the conditions of quiet and moderately disturbed magnetosphere.

It is shown that at each observatory the "tails" of the statistical distributions of impulse amplitudes are described by a power function of the form $f(A) = A^{-\alpha}$ and qualitatively coincide with chaotic regimes peaks intermitted by seldom spikes. An assumption is made that the model of on—off intermittency may be considered as one of the mechanisms of magnetic impulse excitation. It is shown that, depending on local time, season, solar activity cycle phase, geomagnetic latitude of observation, parameters of the solar wind and IMF, the statistical distribution of the MIEs amplitudes follows the regularities typical for the regimes generated in the weakly turbulized medium ($\alpha \sim 1$), or for the chaotic regimes known as "strong turbulence" ($\alpha > 2$). It is assumed that the value of the $\alpha$ index may serve as a characteristic of the level of plasma wave turbulence in the high-latitude magnetosphere.

The obtained results make it possible to assume that: a) the degree of plasma turbulence is larger at higher latitudes of the magnetosphere (77°) than at lower (63°) latitudes; b) the plasma is stronger turbulized in the afternoon sector of the magnetosphere than in the prenoon sector; c) in the Northern Hemisphere of the magnetosphere the plasma is stronger turbulized in summer, whereas in the Southern Hemisphere it is stronger turbulized in winter; d) the plasma is stronger turbulized at the rise phase of solar activity cycle, than at the decay phase. The dependence of the $\alpha$ index on the solar wind speed, longitude $\phi$ of IMF, and the $Bz$ component orientation manifests about the influence of the interplanetary conditions on the plasma of the high-latitude magnetosphere, where MIEs are formed, and also about the changes in the degree of turbulence under the action of external factors.

**Acknowledgments**. The authors thank L. V. Gurieva and S. N. Rumyantseva for their assistance in processing of experimental data. The work was supported by the Russian Foundation for Basic Research (project 03-05-64545).

**References**

Arnoldy, R. L., M. J. Engebretson, J. L. Alford, R. E. Erlandson, and B. J. Anderson, Magnetic impulse events and associated Pc1 bursts at dayside high latitudes, *J. Geophys. Res.*, 101(A4), 7793, 1996.

Bartucelli, M., P. Constantin, C. R. Doering, J. D. Gibbon, and M. Gisselfalt, On the possibility of soft and hard turbulence in the complex Ginzburg-Landau equation, *Physica D.*, 44, 421, 1990.

Bering III, E. A., L. J. Lanzerotti, J. R. Benbrook, Z.-M. Lin, C. G. Maclennan, A. Wolfe, R. E. Lopez, and E. Friis-Christensen, Solar wind properties observed during high-latitude impulsive perturbation events, *Geophys. Res. Lett.*, 17(5), 579, 1990.

Control of the Risk. Stable Development. Sinergetics (in Russian), 431 pp., Nauka, Moscow, 2000.

Dennis, J. E. (Jr), and R. B. Schnabel, Numerical Method for Unconstrained Optimization and Nonlinear Equation (in Russian), 440 pp., Mir, Moscow, 1988.

Glassmeier, K.-H., M. Honisch, and J. Untiedt, Ground-based and Satellite observations of traveling magnetospheric convection twin vortices, *J. Geophys. Res.*, 94(A3), 2520, 1989.

Huber, P. J., Robust Statistics (in Russian), 303 pp., Mir, Moscow, 1984.

Iwasaki, H., and S. Toh, Statistics and structures of strong turbulence in a complex Ginzburg-Landau equation, Progr. Theor. Phys., 87(5), 1127, 1992.

Kataoka, R., H. Fukunishi, and L. J. Lanzerotti, Statistical identification of solar wind origins of magnetic impulse events, *J. Geophys. Res.*, 108(A12), 1436, doi:10.1029/2003JA010202, 2003.

Klain, B. I., N. A. Kurazhkovskaya, About the possible generation mechanism of magnetic impulses events, Abstracts. Scientific Assembly IAGA-IASPEI, Hanoi, Vietnam, 19-31 August 2001, 2001.

Klain, B. I., N. A. Kurazhkovskaya, and O. D. Zotov, Magnetic impulse events and wave turbulence of a high-latitude magnetosphere, European Geosciences Union 1st General Assembly, Nice, France, 25-30 April 2004. Geophysical Research Abstracts, V. 6. SRef-ID: 1607-7962/gra/EGU04-01732, 2004.

Konik, R. M., L. J. Lanzerotti, A. Wolfe, C. G. Maclennan, and D. Venkatesan, Cusp latitude magnetic impulse events. 2. Interplanetary magnetic field and solar wind conditions, *J. Geophys. Res.*, 99(A8), 14,831, 1994.

Konik, R. M., L. J. Lanzerotti, C. G. Maclennan, and A. Wolfe, Cusp latitude magnetic impulse events. 3. Associated low-latitude signatures, *J. Geophys. Res.*, 100(A5), 7731, 1995.

Korotova, G. I., T. J. Rosenberg, L. J. Lanzerotti, and A. T. Weatherwax, Cosmic noise absorption at South Pole Station during magnetic impulse events, *J. Geophys. Res.*, 104(A5), 10,327, 1999.

Lanzerotti, L. J., Conjugate spacecraft and ground-based studies of hydromagnetic phenomenon near the magnetopause, *Adv. Space Res.*, 8, 301, 1989.

Lanzerotti, L. J., R. M. Konik, A. Wolfe, D. Venkatesan, and C. G. Maclennan, Cusp

latitude magnetic impulse events. 1. Occurrence statistics, *J. Geophys. Res*., 96(A8), 14,009, 1991.

Lin, Z. M., E. A. Bering, J. R. Benbrook, B. Liao, L. J. Lanzerotti, C. G. Maclennan, A. N. Wolfe, and E. Friis-Christensen, Statistical studies of impulsive events at high latitudes, *J. Geophys. Res.,* 100(A5), 7553, 1995.

Malinetsky, G. G., and A. B. Potapov, Current Problems in Nonlinear Dynamics (in Russian), 336 pp., Editorial URSS, Moscow, 2000.

Mende, S. B., R. L. Rairden, L. J. Lanzerotti, C. G. Maclennan, Magnetic impulses events and associated optical signatures in the dayside aurora, *Geophys. Res. Lett.*, 17(2), 131, 1990.

Mio, K., T. Ogino, K. Minami, and S. Takeda, Modified nonlinear Schrodinger equation for alfven waves propagating along the magnetic field in cold plasmas, *J. Phys. Soc. Japan*, 41(1), 265, 1976.

Mjolhus, E., On the modulational instability of hydromagnetic waves parallel to the magnetic field, *J. Plasma Physics*, 16(3), 321, 1976.

Moretto, T., D. G. Sibeck, and J. F. Watermann, Occurrence statistics of magnetic impulsive events, *Ann. Geophys*., 22, 585, 2004.

Platt, N., E. A. Spiegel, and C. Tresser, On-off intermittency: a mechanism for bursting, *Phys. Rev. Lett.*, 70, 279, 1993.

Savin, S. P., et al., Turbulent boundary layer at the boundary of a geomagnetic trap, *Letters to JETPh* (in Russian), 74(11), 620, 2001.

Shields, D. W., et al., Multistation studies of the simultaneous impulsive events, *J. Geophys. Res.*, 108(A6), 1225, doi: 10.1029/2002JA009397, 2003.

Sibeck, D. G., and G. I. Korotova, Occurrence patterns for transient magnetic field signatures at high latitudes, *J. Geophys. Res*., 101(A6), 13,413, 1996.

Zakharov, V. E., A. N. Pushkarev, V. F. Shvarts, and V. V. Yan'kov, On soliton turbulence, *Letters to JETPh* (in Russian), 48(2), 79, 1988.

Yahnin, A., E. Titova, A. Lubchich, T. Bosinger, J. Manninen, T. Turunen, T. Hansen, O. Troshichev, and A. Kotikov, Dayside high latitude magnetic impulsive events: their characteristics and relation to sudden impulses, *J. Atm. Terr. Phys.*, 57(13), 1569, 1995.